\documentclass[aps,epsfig]{revtex4} 
\input epsf
\newcommand{\sfig}[2]{
\centerline{ \epsfxsize = #2 \epsfbox{#1} }
		}
\newcommand{\Sfig}[2]{
	\begin{figure}[thbp]
	\sfig{#1.eps}{0.7\columnwidth}
	\caption{{\small #2}}
	\label{fig:#1}
	\end{figure}
}
\newcommand{\Rf}[1]{\ref{fig:#1}}

\def\cmm2{{\,\rm cm^{-2}}}
\def\cm2{{\,{\rm cm}^2}}
\def\cmm3{{\,{\rm cm}^{-3}}}
\def\gcmm3{{\,{\rm g\,cm^{-3}}}}

\def\fun#1#2{\lower3.6pt\vbox{\baselineskip0pt\lineskip.9pt
  \ialign{$\mathsurround=0pt#1\hfil##\hfil$\crcr#2\crcr\sim\crcr}}}

\def\be{\begin{equation}}
\def\ee{\end{equation}}
\def\bea{\begin{eqnarray}}
\def\eea{\end{eqnarray}}
\newcommand{\vs}{\nonumber\\}

\newcommand{\ec}[1]{Eq.~[\ref{eq:#1}]}
\newcommand{\Ec}[1]{[\ref{eq:#1}]}
\newcommand{\eql}[1]{\label{eq:#1}}
\newcommand{\dcl}{\chi_L}
\newcommand{\dcs}{\chi_S}
\newcommand{\rhos}{\rho_s}
\newcommand{\rs}{r_s}
\newcommand{\vm}{v_{\rm max}}

\newcommand{\vc}{v_{\rm cutoff}}

\begin{document}

\title{Galaxy-CMB Lensing} 

\author{Scott Dodelson$^{1,2}$ and Glenn~D.~Starkman$^{3}$
}

\affiliation{$^1$NASA/Fermilab Astrophysics Center
Fermi National Accelerator Laboratory, Batavia, IL~~60510-0500}
\affiliation{$^2$Department of Astronomy \& Astrophysics, The University of Chicago, 
Chicago, IL~~60637-1433}
\affiliation{$^3$Department of Physics, Case Western Reserve University, Cleveland,
OH}

\date{\today}

\begin{abstract}
A long-standing problem in astrophysics is to measure the mass
associated with galaxies. Gravitational lensing provides one of the
cleanest ways to make this measurement. To date, the most powerful
lensing probes of galactic mass have been multiply imaged QSO's (strong
lensing of a background point source) and galaxy-galaxy lensing (weak
deformation of many background galaxies). Here we point out that the mass
associated with galaxies also lenses the Cosmic Microwave Background (CMB)
and this effect is potentially detectable in small scale experiments.
The signal is small (roughly a few tenths of $\mu$K) but has a characteristic
shape and extends out well beyond the visible region of the galaxy.
\end{abstract}
\maketitle


\section{Introduction.}

Flat rotation curves at large distance from galactic centers
imply that the mass associated with a galaxy extends
far beyond the region that is visible. Indeed, it now appears that the vast
majority of the mass of a galaxy is in an unseen component, dark matter.
How much dark matter is there in a galaxy and how far out does the distribution
go? How does the matter associated with a single galaxy compare with the
overdensity around that galaxy due to large scale structure? How is the dark
matter distributed inside the galaxy? Is there substructure, as expected in simulations
of Cold Dark Matter (CDM) models~\cite{moore,andrei,bullock}, or does the absence of many satellites 
for our Galaxy imply that the CDM models need to be
modified\,\cite{spergel,yoshida,colin,burkert}?

These questions increasingly interest not only astrophysicists
studying galaxies and their properties, but also cosmologists and 
particle physicists, for these phenomena may depend on the power spectrum of the
matter in the universe. This in turn depends on such fundamental quantities as 
neutrino masses and the details of inflation. If the substructure dilemma remains,
the properties of dark matter, such as its scattering cross-section or annihilation rate,
may be responsible. These properties may be important clues in identifying
the elementary particles which constitute the dark 
matter and the fundamental theory governing their interactions.

The traditional method of studying the mass distribution in galaxies --
measuring rotation velocities -- has recently been supplemented with a relatively
new technique: studying the deflection of light rays as they pass by or through the
galaxy. There are a number of ways in which gravitational lensing can probe
the mass distribution of an object like a galaxy, but two have emerged recently
as particularly promising. First is the phenomenon of multiply imaged QSOs. Light
from a background point source (the QSO) gets lensed so that multiple images are seen.
The separations between these images, and more importantly their magnifications,
are sensitive to the local mass distribution~\cite{mao,
metcalfe,dalal,chiba,keeton}. Second, images of background
galaxies are distorted by a foreground lensing galaxy, and the amplitude of this
distortion as a function of projected distance from the center of the lens galaxy contains
important information about the mass associated with that galaxy~\cite{tyson,brainerd,fischer,smith,wkl,
mckay, hoekstra,klein}. Due to the presence
of large scale structure, the lensing mass extends far beyond the visible region of the galaxy.
If the diameter of the visible region is of order $10$ kpc, the region within a sphere almost a thousand times
larger will be overdense. 
Understanding this overdensity is an important step
along the way to understanding galaxy formation and the correlation between mass and
light in the universe~\cite{berwei,guzik,cooray,weidav,bwb,jain}.

In this paper, we explore the possibility that
the most distant sources of all, the Cosmic Microwave Background (CMB)
anisotropies generated at redshift $z\simeq 1100$, can serve as the sources
which are lensed by foreground galaxies.
The problem is very similar to the lensing of the CMB by foreground
galaxy clusters~\cite{sz}, except that the amplitude of the signal
is considerably smaller here.
The observed temperature $T$ as a function of 2D position $\vec\theta$
on the sky is
\bea
T(\vec\theta) &=& \tilde T(\vec\theta') = \tilde T(\vec\theta-\vec{\delta\theta})
\vs
&\simeq & \tilde T(\vec\theta) - \vec{\delta\theta}\cdot \nabla \tilde
T(\vec\theta)
\eea
where $\tilde T$ is the background CMB field originating from the unlensed
position $\vec\theta'=\vec\theta -\vec{\delta\theta}$, and $\delta\theta$ is the deflection
due to the lens. As illustrated in \cite{sz}, 
the CMB anisotropy field on small scales is almost purely a dipole (or gradient; on these small scales the
two terms are interchangeable);
aligning the $y-$ axis
with the dipole leads to 
a simple expression for the observed temperature:
\be
T(\vec\theta) \simeq {\partial\tilde T\over \partial \theta_y} \times \left( \theta_y - \delta\theta_y\right)
.\eql{tbasic}\ee
Here $\partial T/\partial\theta_y$ is the temperature gradient, which is of course zero on average. Its
rms fluctuation though is equal to $0.22\mu K$ arcsec$^{-1}$ in the standard $\Lambda$CDM
cosmology with $\Omega_{\rm matter}=0.3$, $\Omega_{\rm baryon} = 0.04$, $H_0=70$ km
\,sec$^{-1}$\,Mpc$^{-1}$
and $\Omega_\Lambda=0.7$ (the rest of this paper assumes this background
cosmology). The deviations 
from a pure dipole arise from the deflection angle
$\vec{\delta\theta}$. We will see that the deflection angle induced by
the mass associated with a galaxy is typically of order an arcsecond, so the
expected signal is small, tenths of $\mu$K per galaxy. Eq.~\Ec{tbasic} breaks down
on scales larger than ten arcminutes when the background CMB no longer behaves
as a pure gradient due to the power on the scales of the acoustic peaks~\cite{mc}. 
Even on smaller scales, the ``noise'' due to the quadrupole can be significant
as we will see in \S{III}.

\S{II} introduces the basic formula relating
the deflection angle to the mass distribution and presents the deflection
angle for several simple mass distributions. 
\S{III} considers the lensing due to the large scale overdensities
surrounding a galaxy. These extend out to several Mpc (or tens of arcminutes
for a galaxy at redshift $0.1$). Given the resolution and sensitivity of
current CMB experiments, this large scale lensing is likely to be
the easiest target for future experiments. Just as in the galaxy-galaxy lensing
measurements, the signal from a single foreground galaxy is quite small, so 
the statistics of many galaxies must be used to beat down the noise.
\S{IV} discusses the foregrounds that might contaminate this signal.
We let our fancy take flight in \S{V} where we consider the signal
due to a single galaxy and show that in principle the lensed CMB can detect
the sub-structure that simulations predict must exist in galactic halos. Observing
this signal would require experiments with significantly better resolution and
sensitivity than those currently planned, but if the small scale problems of CDM persist,
we should keep this technique in mind as a way of definitively resolving
the ``small scale crisis.'' 

One final comment about previous work.
Several years ago, Peiris and Spergel\,\cite{peiris} analyzed many aspects of the CMB-galaxy
cross-correlations. Their analysis, focused on the WMAP experiment\,\cite{wmap},
and therefore is for scales larger than we consider here. 

\section{Deflection Angle and Mass Models}

The deflection angle of a photon due to a galaxy at redshift $z_L$,
at comoving distance 
from us, $\dcl$, is\,\cite{mcex}
\be
\vec{\delta\theta} = -8\pi G {\vec\theta\over\theta}\, 
 {(\dcs-\dcl) \over \dcs}\,
\left[
{1+z_L\over \theta\dcl}\int_0^{\dcl\theta/(1+z_L)} dR\, R ~\Sigma(R) 
\right]
.\eql{dtheta}\ee
Here we take the photon to be emitted from a comoving distance 
$\dcs=1.4\times 10^4$ Mpc away from us, corresponding to the surface of
last scattering. Its unperturbed path has at all times an angular distance $\theta$
from the axis connecting us to the center of the galaxy. The integral
is over the two dimensional mass density $\Sigma$ of the galaxy, which is
assumed to be azimuthally symmetric.

A singular isothermal sphere has a density profile proportional to $r^{-2}$, so
that the surface density is proportional to $R^{-1}$, where $r\, (R)$ is the 
3D (2D) radius from the center of the galaxy. For this profile then the term in
square brackets in \ec{dtheta} is independent of $\theta$ and the amplitude is
fixed by the correlation length $r_0$ defined via
\be
\rho(r) = \rho_m (r_0/r)^2
.\ee
Here $\rho_m$ is the average matter density in the universe, so that $8\pi G \rho_m = 0.9 H_0^2$
with $c/H_0 = 4300$ Mpc. The surface density for this distribution is $\Sigma(R)=\pi \rho_m r_o^2/R$,
so the amplitude of the deflection angle due to
a singular isothermal sphere is
$0.9 H_0^2 \pi r_0^2 = 1'' (r_0/5.6\, {\rm Mpc})^2$ for $\dcs\gg \dcl$. Simulations~\cite{guzik}
suggest that the distribution of matter around galaxies, which is formally measured
by the galaxy-mass cross-correlation function may indeed be represented by
this distribution with $r_0\simeq 7$ Mpc. The distortion, at least on scales
larger than about $100$ kpc, is therefore expected to be constant with an amplitude of a little
more than an arcsecond. Figure~\Rf{rhoint} shows this deflection angle; it falls off 
on large scales since we do not include contributions from the density beyond $r_0$.
Also shown is a more conservative choice of $\rho=\rho_m (r_0/r)^{-1.8}.$
Even for this choice, the expected deflection angle is of order an arcsecond. 

It might seem at first that detecting such a small deflection will be hopeless
given that the rms deflection of a photon from the CMB is as much as 3 arcminutes.
However, this is due almost entirely to lensing by large scale (low $\ell$) fluctuations.
Large patches of the CMB sky will thus be nearly uniformly  deflected by up to several
arcminutes, however, by $\ell=10^4$, the rms deflection is down to about half an 
arc-second.   Thus, CMB lensing by galaxies will take place against a relatively
smoothly-lensed background, which is still well described by a local dipole.  

\Sfig{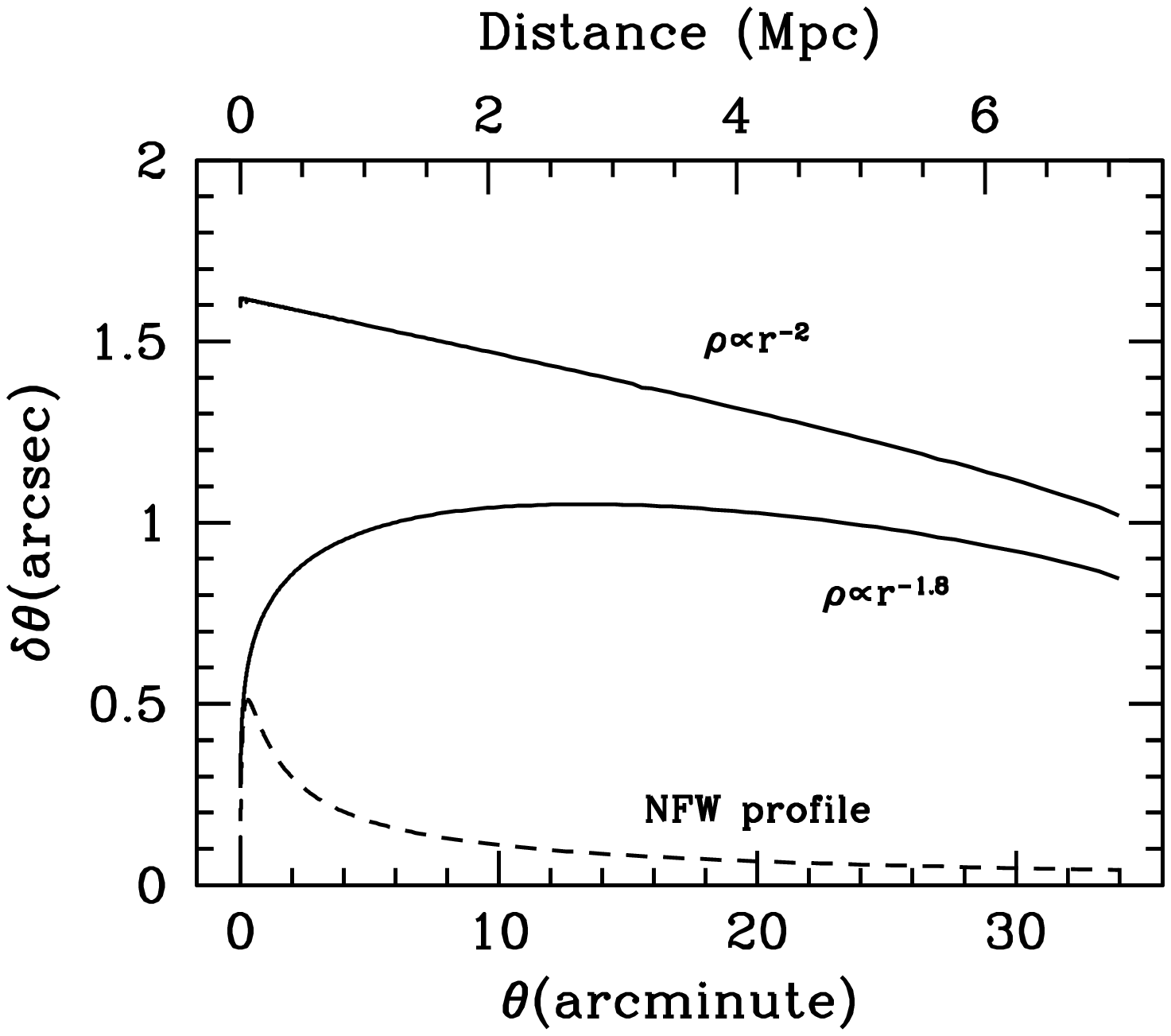}{The deflection angle as a function of distance from the galactic center
for several different mass models. The top two curves have an overdensity equal to
$\rho_m \xi(r)$ with the galaxy-mass cross-correlation function equal to $(7 {\rm Mpc}/r)^\alpha$
and $\alpha=1.8$ and $2$. The bottom curve is due to an individual halo with 
an NFW profile with concentration $c=8$ and maximum rotational velocity equal to
$200$ km sec$^{-1}$. The galaxy is assumed to be at redshift $z_L=0.2$ in order to
obtain the relation between angular distance (bottom axis) and physical distance (top
axis).}

The above estimates are for a statistical sample of galaxies, accounting for the
fact that the universe is clustered, so we expect halos of dark matter to
reside near each other. These estimates definitely break down on scales smaller than 
$\sim 100$ kpc. On these smaller scales, a more appropriate distribution is
given by the Navarro, Frenk, and White (NFW)~\cite{nfw}
profile:
\be
\rho(x\equiv r/\rs) = {\rhos \over x(1+x)^2}
.\ee
The NFW profile can be described by two parameters: the scale radius
$\rs$ and the density parameter $\rhos$. More commonly, these are traded in for
the concentration $c$, which is the ratio of the virial radius to the scale
radius, and $\vm$, the maximum rotational velocity due to this mass distribution.
The virial radius is defined as that within which the average density
is equal to $\Delta\rho_m$ with $\Delta=337$\,\cite{bryan,bul337} for the $\Lambda$CDM model 
in which we are working. The maximum rotational velocity can be
determined analytically in terms of $\rs$ and $\rhos$; it is
$\vm = 0.46(4\pi G\rhos\rs^2)^{1/2}$. Figure~\Rf{rhoint} shows the
deflection due to an NFW profile with $\vm=200$ km sec$^{-1}$
and concentration $c=8$. Analytically,
\be
\delta\vec\theta\Big|_{\rm NFW} = 
-0.54''\, {\vec\theta\over\theta} \, \left( { \vm\over 200 {\rm km\ sec}^{-1}} \right)^2
  {(\dcs-\dcl)\over \dcs} f(\lambda) ,
\eql{nfwlens}  \ee
  where $f(\lambda)$ is a smooth function which reaches its peak of one at
  $\lambda=1.3$:
  \be
  f(\lambda) \equiv 
\frac{3.23}{\lambda}\cases{ \ln(\lambda/2) + {\ln(\lambda/[1-\sqrt{1-\lambda^2}])\over
\sqrt{1-\lambda^2} } & $\lambda < 1$
\cr
\ln(\lambda/2) + {\pi/2 -\arcsin(1/\lambda) \over \sqrt{\lambda^2-1} } 
& $\lambda > 1$.}
\ee
$\lambda$ is defined as
\be
\lambda \equiv {\dcl\theta\over (1+z_L) \rs}
.\ee

\section{Large Scale Galaxy-CMB Lensing}

Recently, the Sloan Digital Sky Survey (SDSS) detected the signal
of galaxy-galaxy lensing out to a Mpc~\cite{mckay}. It is useful to
recap their technique to place galaxy-CMB lensing in context.
They select lenses as those galaxies bright enough for spectra to have
been taken (Petrosian magnitude $r^*<17.6$) and background galaxies
fainter than $18$th magnitude. For the background galaxies, photometric
redshifts are used. About $30,000$ lens galaxies are chosen from SDSS
commissioning data and these are probed by about $3\times 10^6$ background
galaxies. Thus each lens has about a hundred background galaxies behind it.
The error in the shear produced by a single foreground galaxy is $0.4/\sqrt{N_{bg}}$
where the $0.4$ comes from a combination of instrumental noise and the
intrinsic ellipticities of the background galaxies.
Thus, if $N_{bg}=100$, the noise in the shear is $0.04$. This is almost a factor of
ten larger than the expected signal. So the SDSS survey cannot detect galaxy-galaxy
lensing due to a single galaxy. Instead they must average the signal over many foreground galaxies
(each with a signal to noise of order $0.1$). SDSS nevertheless provides a useful measure of the
galaxy-mass correlation function because there are many ($3\times 10^4$) foreground galaxies
over which to average.

How does the SDSS signal to noise compare to that obtainable with the CMB?
We have determined (Eqs.~\Ec{tbasic} and \Ec{dtheta}) that the expected signal 
in a CMB experiment is
\be
T(\vec\theta) - {\partial T\over\partial\theta_y} \theta\cos\phi ={\partial T\over\partial\theta_y}  b(\theta) \cos\phi
\ee
where $\phi$ is the angle between the angular position $\vec\theta$ and the $y$-axis along
which the background gradient is oriented and $b(\theta)$ is roughly constant with
an amplitude of order one arcsecond. We also need to consider the deviation of the background anisotropy pattern $\tilde T$
from a pure gradient.
On scales larger than $5-10'$, the background field is more complicated. For simplicity we will
assume that the background dipole can be removed but the quadrupole cannot, and it serves
as a source of noise. That is, we consider the next term in the expansion of $\tilde T$:
\be
\tilde T(\vec\theta) \simeq 
{\partial T\over\partial\theta_y} \theta\cos\phi + {1\over 2} \theta_i\theta_j {\partial^2\tilde T\over \partial \theta_i \partial \theta_j}
.\ee
This second term on the right also has mean of zero, but variance equal to 
\be
\sigma_Q^2 = {3\over 32} \theta^4 \sum_l {l^5 C_l\over 2\pi}
= (1.5 \mu K)^2 \left( {\theta\over 1'} \right)^4
\ee
where the second equality holds in the standard cosmology we are considering. If the quadrupole cannot
be removed, then this source of
noise adds in quadrature with that due to the atmosphere and/or instrument. 

We can now compute the signal to noise expected for lensing of the CMB 
due to a single galaxy. Consider a CMB experiment which maps the sky
into $N_p$ pixels each of area $\Delta\Omega$ with instrumental/atmospheric noise per pixel $\sigma_n$. 
Then,
\bea
\left( {S\over N} \right)^2 &=& \langle\left( {\partial\tilde T\over\partial\theta_y} \right)^2 
\rangle\sum_{i}^{N_p} {\cos^2\phi_i\over 
\sigma_n^2 + \sigma_Q^2(\theta_i)} b^2(\theta_i)\cr
&\simeq &
\langle\left( {\partial\tilde T\over\partial\theta_y} b\right)^2 \rangle\int {d^2\theta\over \Delta\Omega} {\cos^2\phi\over 
\sigma_n^2 + \sigma_Q^2(\theta)} = \left[ 0.089 \Big( {1'\over \sqrt{\Delta\Omega}} \Big) \Big( {10 \mu K \over \sigma_n} 
\Big)^{1/2}\right]^2
.\eql{sn}\eea
So, for a CMB experiment to achieve resolution comparable to SDSS (signal to noise
per galaxy of $0.1$), beams of order an arcminute
with noise per pixel of order $10\mu$K are needed. This is within the range 
expected of upcoming experiments. A note of caution: the $\simeq$ sign in \ec{sn}
is a warning that this signal to noise estimate assumes the pixel size is significantly smaller
than the scale at which the background dipole approximation breaks down; i.e.
$\Delta\Omega \ll (10')^2$.

One might wonder whether, given the functional dependance of $\delta\vec\theta$ on the lens
distance/redshift (\ec{nfwlens} and Ref.~\cite{song} for example), it might not be more advantageous to look to higher redshift 
objects rather than SDSS galaxies as lenses  For example, lyman-alpha/star-forming regions at $z\simeq 3$ 
might be expected to give significantly higher a $S/N$.  Studying the mass distribution in such objects
would indeed be of significant interest; however, since these high redshift objects will
subtend  a significantly smaller solid angle than medium-to-low redshift galaxies, it
will have to wait until a dramatic improvement in angular resolution before their
CMB-lensing signal  can be properly studied.  We therefore focus our attention on SDSS-type
galaxies as lenses. 

Since the noise from the quadrupole dominates over instrumental
noise when $\theta> 1' (\sigma_n/1.5\mu K)^{1/2}$, and a fair fraction of the signal comes from these larger 
scales, reducing instrumental noise is not as a important as going to higher resolution and/or covering
more sky. That is, signal to noise estimates typically scale as $(\Delta\Omega \sigma_n^2)^{-1/2}$ 
when the dominant noise is instrumental. Here, though, the scaling is $(\Delta\Omega \sigma_n)^{-1/2}$. 
For fixed resolution, then, the final signal to noise scales as $N_{rm gal}^{1/2} \sigma_n^{-1/2} 
\propto N_p^{1/2} \sigma_n^{-1/2}$. The number of pixels covered if the total time is fixed is inversely
proportional to the time spent on each individual pixel, while $\sigma_n$ is inversely proportional to 
the square root of time per pixel. Therefore, the final signal to noise scales increases as less time is
spent on each pixel: an experiment intent on measuring CMB-galaxy lensing should
strive for large sky coverage at the expense of sensitivity.

For CMB experiments with larger beams, the approximation that the background source
is a gradient breaks down. Information is still contained in the cross-correlation of
the CMB and galaxy surveys~\cite{peiris}. As we have seen the signal is of order a tenth of a 
$\mu$K, while the noise per pixel due to primordial CMB fluctuations is naively of order $50\mu$K. 
Without a sophisticated algorithm to extract the signal, we would then need of order
$500^2$ galaxies to beat down the noise. Going to larger redshifts to
pick up these galaxies would not necessarily help since the galaxy-dark matter cross-correlation
functions falls off
at high redshifts, so the signal is significantly smaller. Also, a galaxy projects to a smaller
angular size at high redshift, necessitating even higher resolution.
Although we do not pursue it here, two possible approaches are: (i) to assume
the form of the lensing profile and fit for its amplitude~\cite{sz,peiris} or (ii)
to extend the likelihood technique developed by Hirata and Seljak~\cite{hirata}
to this cross-correlation case.

\section{Noise}

Before we assert that the lensing signal from galaxies is observable,
we must identify the other potential sources of noise.  
Extensive discussions of CMB noise sources on small angular scales are to be found in 
\cite{tegmark,toffolatti,teho}, and are reviewed in \cite{sz} in the context of cluster lensing,
where the issues are similar.  

The expected sources of noise (other than detector noise and intrinsic noise from
the CMB itself, both considered in \S{III}) are: 
Galactic synchrotron, free-free and dust emission, 
thermal Sunyaev-Zeldovich (SZ) from galaxy clusters and filaments,
kinetic SZ (including Ostriker-Vishniac) resulting from bulk gas motions,
and unresolved point sources.  
We briefly review each of these in turn.
 
\subsubsection{Galactic emissions}
Galactic synchrotron and free-free emission both decline with increasing $\ell$.
Fluctuations due to free-free emission are already below  $0.1\mu$K by $\ell=10^4$
above $30$ GHz \cite{teho}.  Above $100$ GHz synchrotron emission fluctuations are below
$0.1\mu$K by $\ell=10^3$; at 30GHz they are predicted to be $\sim2\mu$K at $\ell\simeq 1000$,
and declining as approximately $\ell^{1/3}$ \cite{tegmark}.  This puts them below $1\mu$K below
one arcminute. Temperature fluctuations due to Galactic dust emissions are expected to be below $1\mu$K
at frequencies below 217 GHz for $\ell > 3000$ if the dust emission is vibrational.
The situation was less clear if the dust emission is rotational \cite{teho}, but the
recent WMAP\,\cite{wmapfor} constraints suggest that rotational dust will not be a problem.

\subsubsection{Thermal SZ}

Of particular interest for galaxies being observed in the foreground of an SZ cluster,  
are fluctuations in the cluster SZ. 
Clusters will themselves have substructure -- component galaxies, fluctuations  in electron
temperature and density across the cluster.  
To the extent that these are resolvable, whether in the CMB data itself, in cluster radio maps,
or in X-ray maps (especially with future X-ray interferometers), they could be subtracted and
one could select as target galaxies those that do not have significant cluster structure
in their backgrounds.  

The unmodeled thermal SZ foreground has been calculated by a number of groups\,\cite{bond,zhang,hernquist}.  
The power is proportional to $\sigma_8^7$, so there remains significant uncertainty in
the amplitude. If $\sigma_8$ is set to its WMAP value of $0.8$\,\cite{wmapspe}, then the expected
rms amplitude of the thermal SZ on the scales of interest is of order $5\mu$K.
Of course, the signal (or in this case, the noise) goes away at $217$GHz
so it is always possible in principle to avoid this source of contamination.  
(The relatvistic correction which spoils the null can be expected to be 
severely depresssed below the $5\mu$K level.)

\subsubsection{kinetic SZ and Ostriker-Vishniac}

The kinetic SZ effect is the Doppler shift imprinted on CMB photons when they scatter
off electrons in a moving gas.  In the linear regime for the matter fluctuations
this is called the Ostriker-Vishniac (OV) effect \cite{vishniac}. 
The kinetic effect has the same spectral shape as the primordial CMB
(and the lensing signal we are after) so it is more difficult to eliminate than the
thermal effect. The amplitude of the effect though is of order $2 \mu$K\,\cite{hu,sil1,sil2,gnedin,hernquist,
val,ma,zhang,zh2}, so it is not
a major stumbling block.

\subsubsection{Point Sources}

Toffolatti {\it et. al.} \cite{toffolatti} considered the contribution of
unresolved point sources to fluctuations in the anisotropy of the CMB on 
1-100 arcminute scales, at the 8 Planck wavebands from 30GHz to 857GHz.
These sources include, in the low frequency channels, radio loud AGNs, 
flat-spectrum radio-galaxies, quasars, and high-z BL-Lacs.
Dust emission from distant dust-rich young galaxies  
is the dominant source at higher frequencies.  
Assuming that point sources with flux below 1mJy remain unsubtracted,
they find (figures 5 and 6) that in all channels $\langle(\Delta T/T)^2\rangle^{1/2} > 10^{-6}$ 
at 1 arcminute (best channel -- 143 GHz), and in many cases is much bigger.  
It is also increasing with  decreasing $\theta$ as approximately $\theta^{-1}$,
suggesting that in the best channels (between about 44 and 150 GHz) we can expect 
point sources to contribute about 5-10 microK of noise at 6 arcsecond , 
and maybe 10-15 microK of noise at 1 arcsecond.

To make much improvement in these noise figures 
you would need to be able to accurately subtract 
very low flux point sources -- about 0.1 mJy to achieve a a factor of 3, 
0.01mJy to get down an order of magnitude.  It is also true that point sources
will not exhibit the coherent extended structure expected from lensing
and may possibly be filtered out.  The biggest confusion is likely to arise
if we attempt to use galaxy lensing to measure the proto-galactic substructures.
On this scale the foreground dust emission is dominant.  However, it also has 
a rather different energy spectrum than the SZ-distorted  CMB, which may aid in
its subtraction.

\subsubsection{Other Sources of Confusion} 

There are several other effects that produce signals similar to that considered
here. Moving clusters produce nonlinear corrections to the Integrated Sachs Wolfe Effect on small scales, and
the pattern produced has the same structure as that due to lensing. The signal
is quite small though\cite{aghanim,coo2} ($\sim 0.3\mu$K) and on larger scales than interest us.
Coherent electron rotational velocities in clusters lead to a dipolar
signature\cite{cooche,chluba} and these are on relatively small scales. But the amplitude
is also very small and the induced signal is not aligned with the cosmic dipole.

\section{Single Galaxy Lensing}

The lensing signal due to the halo surrounding a single galaxy is given
by \ec{nfwlens}. Since the signal is very small, we cannot hope to measure
it with the current, or even currently planned, CMB experiments.
Here we simply motivate future experiments by illustrating that the potential
pay-off is large: lensing of the CMB in principle allows us to differentiate between
an NFW profile and an isothermal profile, thereby testing one result of numerical
simulations. More ambitiously, very high resolution
measurements could detect the sub-structure
predicted by current theories of structure formation.

Detecting the shape of the dark matter profile around a single galaxy
with CMB lensing requires resolution and sensitivity far beyond current
capabilities. Figure~\Rf{med} shows the different signals induced by an NFW profile
and an isothermal profile. The galaxy is situated at $z=0.1$ (the lower the redshift,
the larger it appears, and therefore the less difficult is the issue of resolution).
A number of groups~\cite{bul337} have found that typical values of the concentration
parameter of galaxies at redshift one are $5-10$ in CDM models. 
The galaxy doing the lensing in Figure~\Rf{med} has concentration
parameter equal to eight and a maximum circular velocity of $200$
km sec$^{-1}$. This corresponds to a scale radius $\rs=44$ kpc and a virial mass of 
$2\times 10^{12} h^{-1} M_\odot$. 

\Sfig{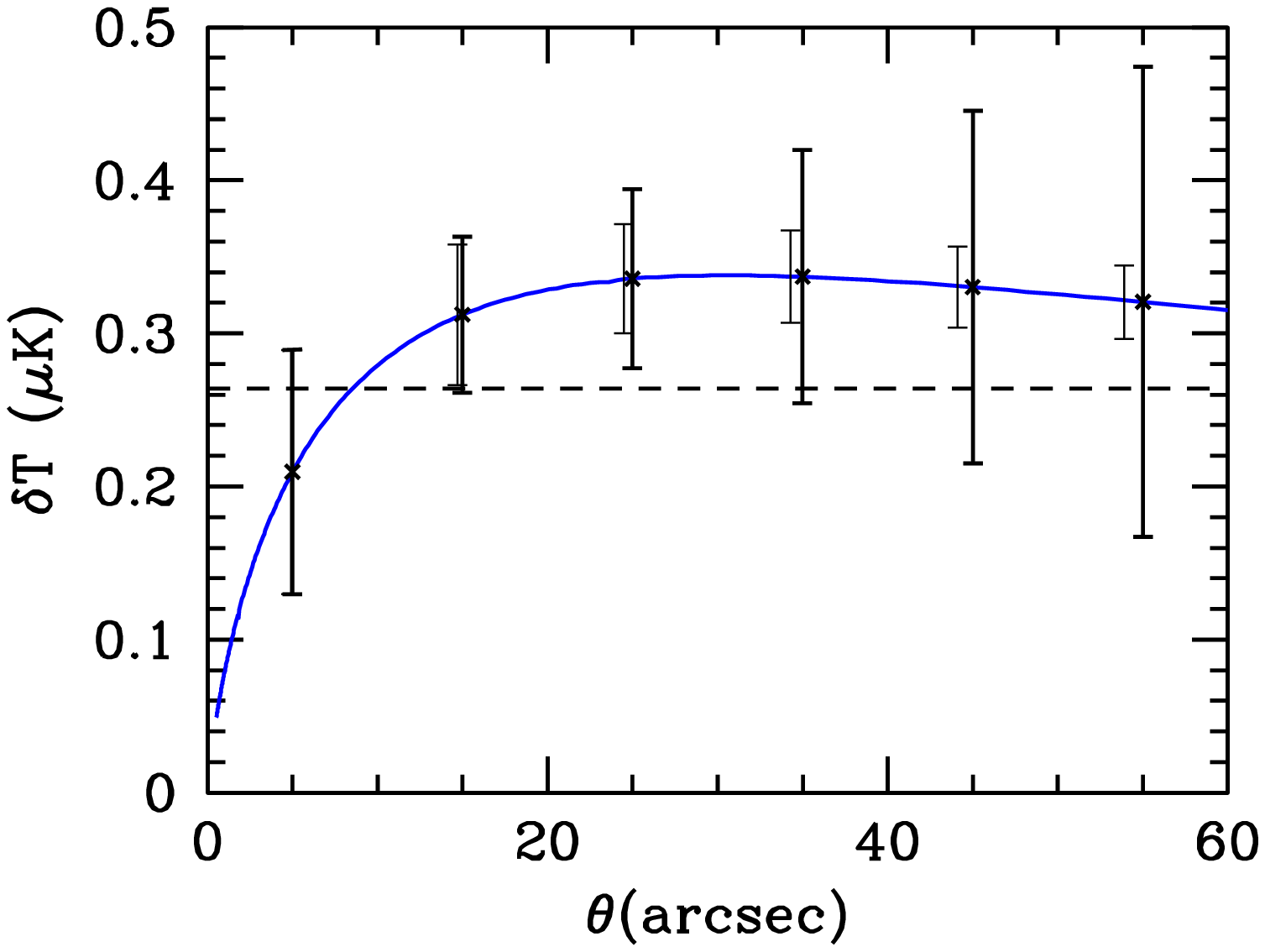}{Anisotropy induced by lensing by an NFW dark matter profile (solid curve)
and an isothermal sphere (dashed straight line), with the galaxy sitting
at redshift $0.1$. The galaxy is chosen to have concentration $8$ and scale radius
equal to $44$ kpc (corresponding to maximum rotational velocity of order $200$ km sec$^{-1}$).
Errors are shown for a $5''$ beam with $0.4\mu$K instrumental noise per pixel. Light
error bars do not include noise from the background quadrupole; larger, heavier bars do.}

Even Figure~\Rf{med}, with its optimistic noise and resolution parameters, paints too
pretty a picture. For, we have neglected all foregrounds, save for the local quadrupole of the CMB. 
Noise due to the quadrupole kicks in at surprisingly small scales, because of the exquisite
sensitivity required to discriminate these models.

The surface density of a smooth NFW profile with these parameters is depicted in 
Figure~\Rf{sig}. Unhindered by experimental complications, we are led to ask
whether a smooth NFW profile could be distinguished from the clumpy profiles found
in simulations.

\begin{figure}[thp]
\centerline{
\hbox{ \hspace{0.0in}
    \epsfysize=3in
    \epsffile{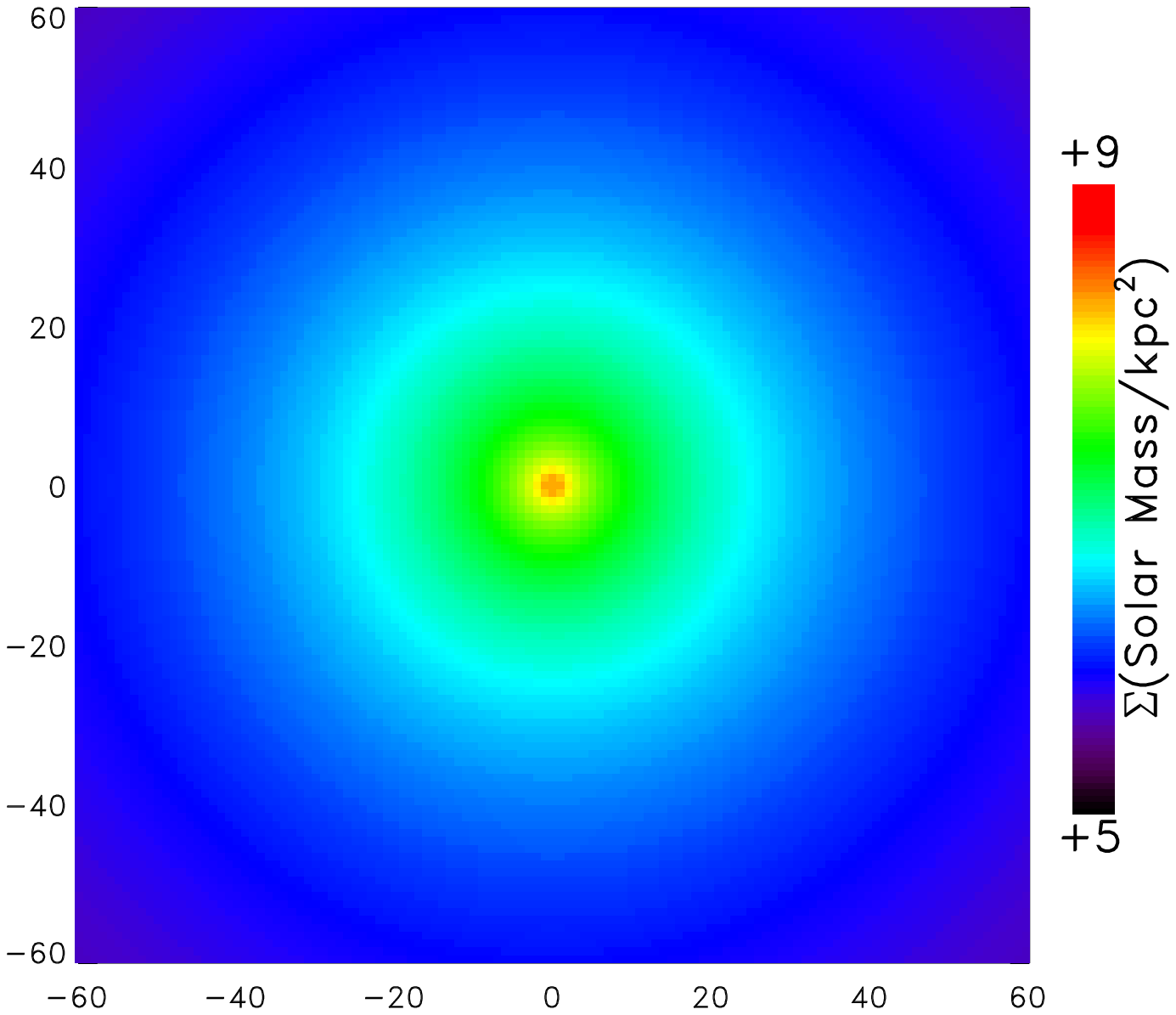}
    \hspace{0.05in}
    \epsfysize=3in
    \epsffile{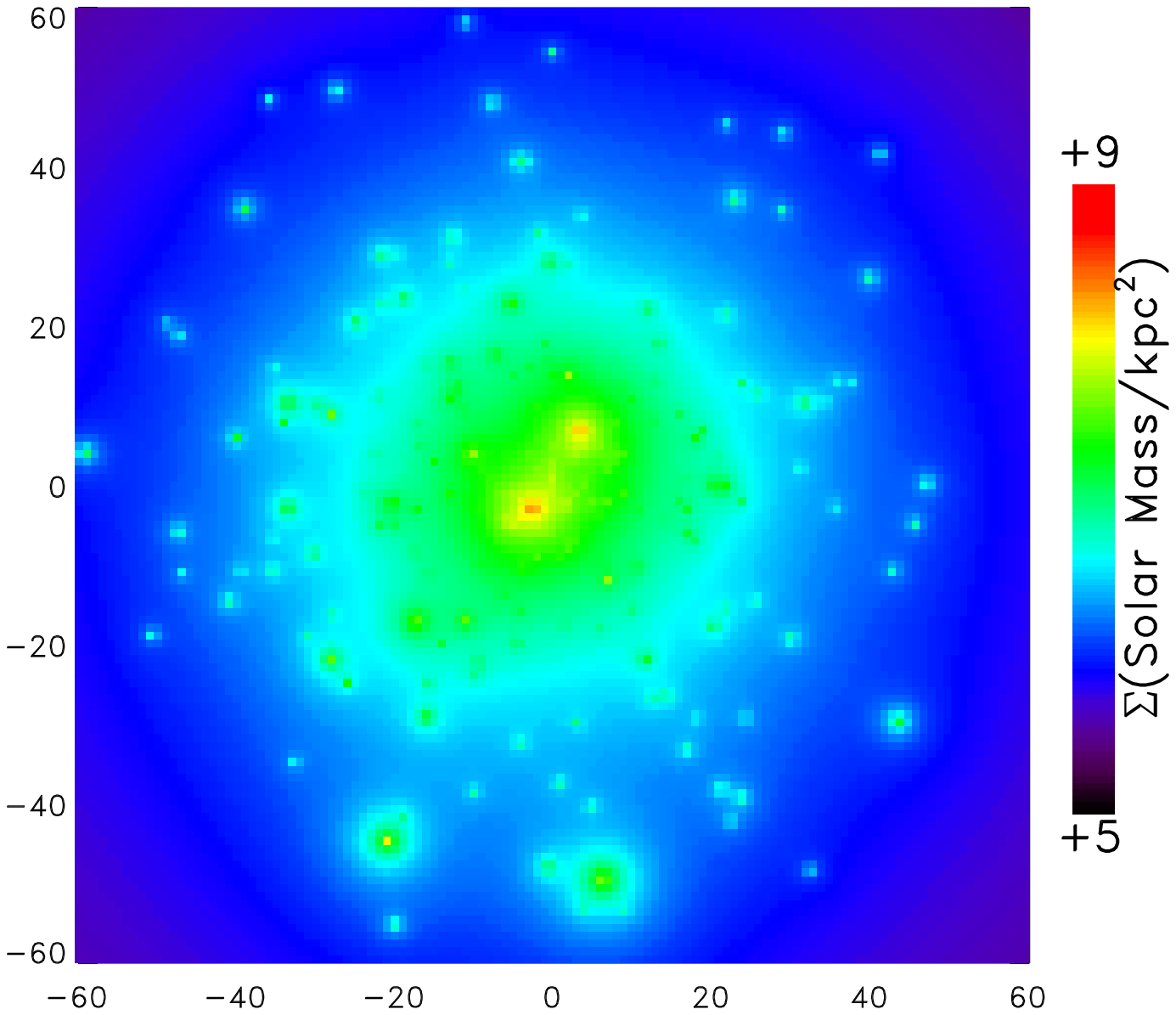}
    }
  \vspace{9pt}
  }
\caption{Log of the surface density of a galaxy at redshift one. 
Axes are in units of arcsec.
{\it Left Panel:} NFW profile with
a scale radius equal to $44$ km and a maximum circular velocity of
$200$ km sec $^{-1}$. {\it Right Panel:} The same mass
distributed in $240$ sub-halos.}
\label{fig:sig}
\end{figure}

To construct a clumpy halo, we use the distribution of sub-halos measured by \cite{moore}. 
They find
that the cumulative number of halos with circular velocity greater 
than $\vm$ is roughly
equal to $0.03 (\vm/v_{\rm parent})^{-3}$. This means that the distribution of
halos 
\be
dn/d\vm \simeq \cases{
0.09 \left( {v_{\rm parent}^3\over \vm^4} \right) & $\vm > \vc
 $\cr
	0 & otherwise}
.\eql{dndv}\ee
 Here the lower cut-off arises because we don't know what the simulations
 predict on such small scales. Note that with
 this distribution, if we take $\vc = 0.05v_{\rm parent}$,
 there are $240$ sub-halos in the galaxy.
 
 To generate a distribution of velocities of sub-halos, first normalize \ec{dndv}
 to unity: $dn^{\rm norm}/d\vm = (1/N) dn/d\vm$ and then given a random
 number $\eta$ between zero and one, define the velocity of one sub-halo $v$ via
 \be
 \eta = \int_{\vc}^v d\vm {dn^{\rm norm}\over d\vm}.\ee
In this case we can do the integral analytically, so 
\be
\vm = { \vc \over (1 - \eta)^{1/3}}
.\ee

We can use the same technique to generate the spatial distribution of halos. 
Suppose that the number of halos is distributed according to
an NFW profile. We want
$\Sigma(R)$ from the sub-structure to follow an NFW profile. The total mass
within a radius $R$ is $2\pi\int_0^R dR'\, R'\,\Sigma_{\rm NFW}(R')$. So the
fraction of the total mass which is within $R$ is
\be
\eta(R) = { \int_0^{R/\rs} dx\,x\,\Sigma_{\rm NFW}(x) \over \int_0^{R_{max}/\rs}
dx\,x\,\Sigma_{\rm NFW}(x) }
.\ee
Here we have cut off the distribution at $R_{\rm max}= 10\rs$; note that $\eta$
lies between zero and one, so a random number between zero and one can be
associated with a given value of $R$. The angular coordinate of the sub-halo is
also assigned randomly. This distribution is shown in the right panel of Figure~\Rf{sig}.

\begin{figure}[thp]
\centerline{\hbox{ \hspace{0.0in}
    \epsfysize=3in
    \epsffile{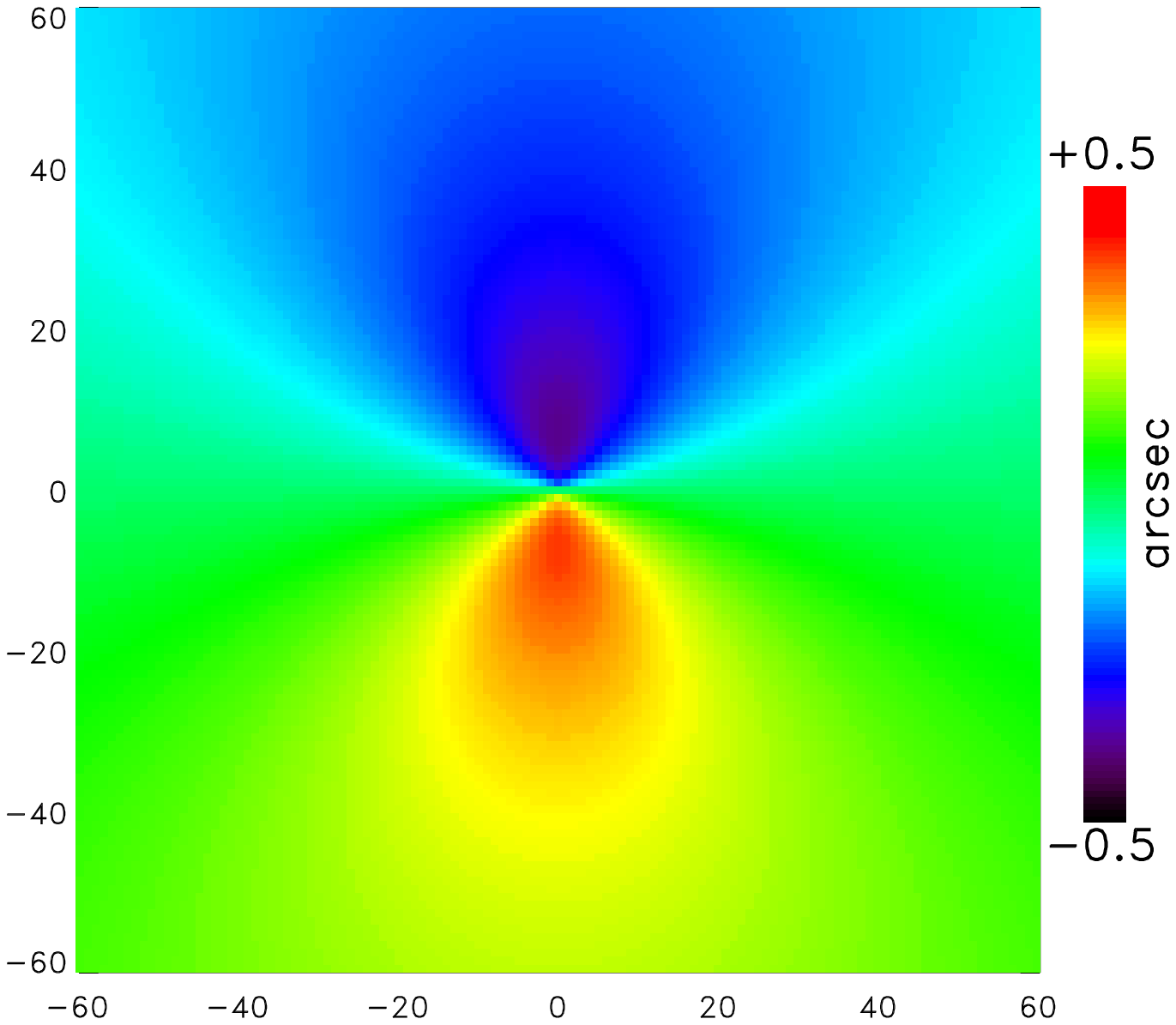}
    \hspace{0.05in}
    \epsfysize=3in
    \epsffile{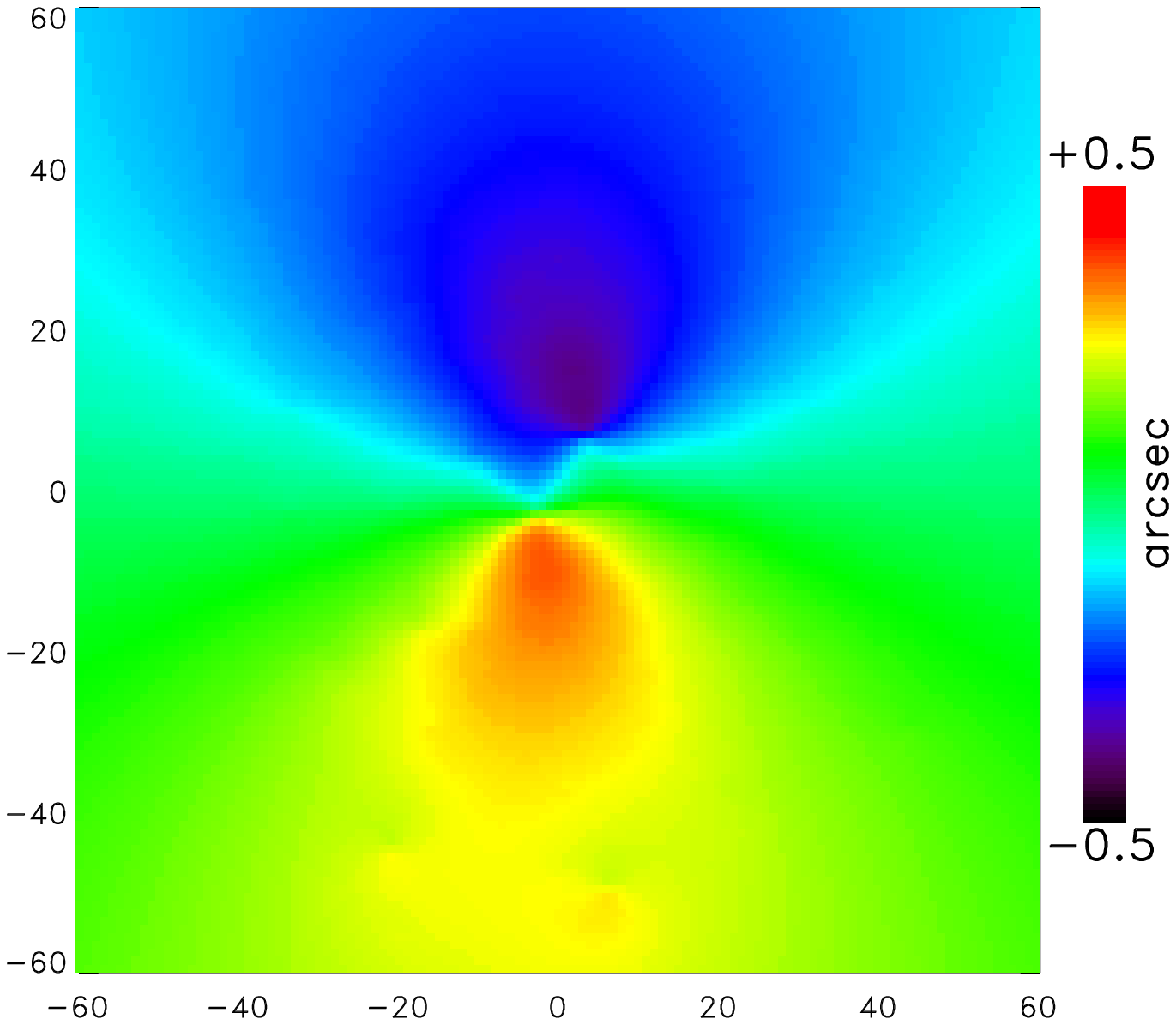}
    }
  }
  \vspace{9pt}
\caption{Temperature profile due to lensing of the background CMB
by a foreground galaxy at redshift one. The background CMB is
a pure dipole aligned from bottom to top.
{\it Left Panel:} Lensing due to smooth galaxy depicted in
left panel of Figure~\Rf{sig}. {\it Right Panel:} Lensing due
to the clumpy galaxy shown in right panel of Figure~\Rf{sig}.}
\label{fig:txty}
\end{figure}

Figure~\Rf{txty} shows the imprint on the CMB resulting
from both the smooth and the clumpy galaxy. Note the characteristic
signal first pointed out in Ref.~\cite{sz}: a hot and cold lobe on
either side of the galactic center. (Here we have subtracted off the
dipole.) 
Traces of the substructure are still evident in the right panel
of Figure~\Rf{txty}. However, these are much less obvious than those
in the surface density plot of Figure~\Rf{sig}. The similarity of
the two panels in Figure~\Rf{txty} is reflection of the fact that 
the galaxy is lensing a large ``sheet,'' the background dipole, as
opposed to the point-like QSO's with which we are more familiar~\cite{pkeet}.

\begin{figure}[thp]
\centerline{\hbox{ \hspace{0.0in}
    \epsfysize=3in
    \epsffile{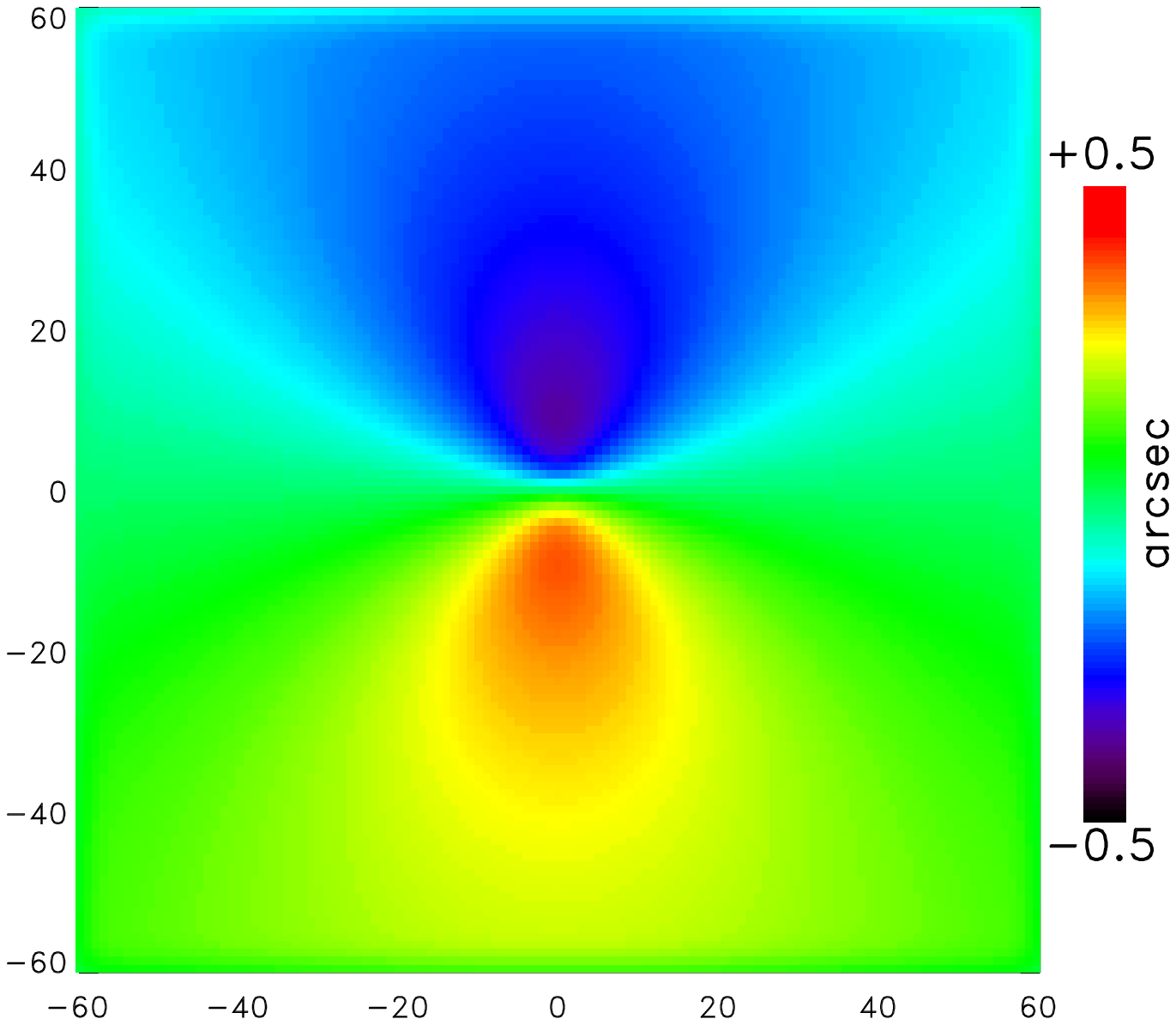}
    \hspace{0.05in}
    \epsfysize=3in
    \epsffile{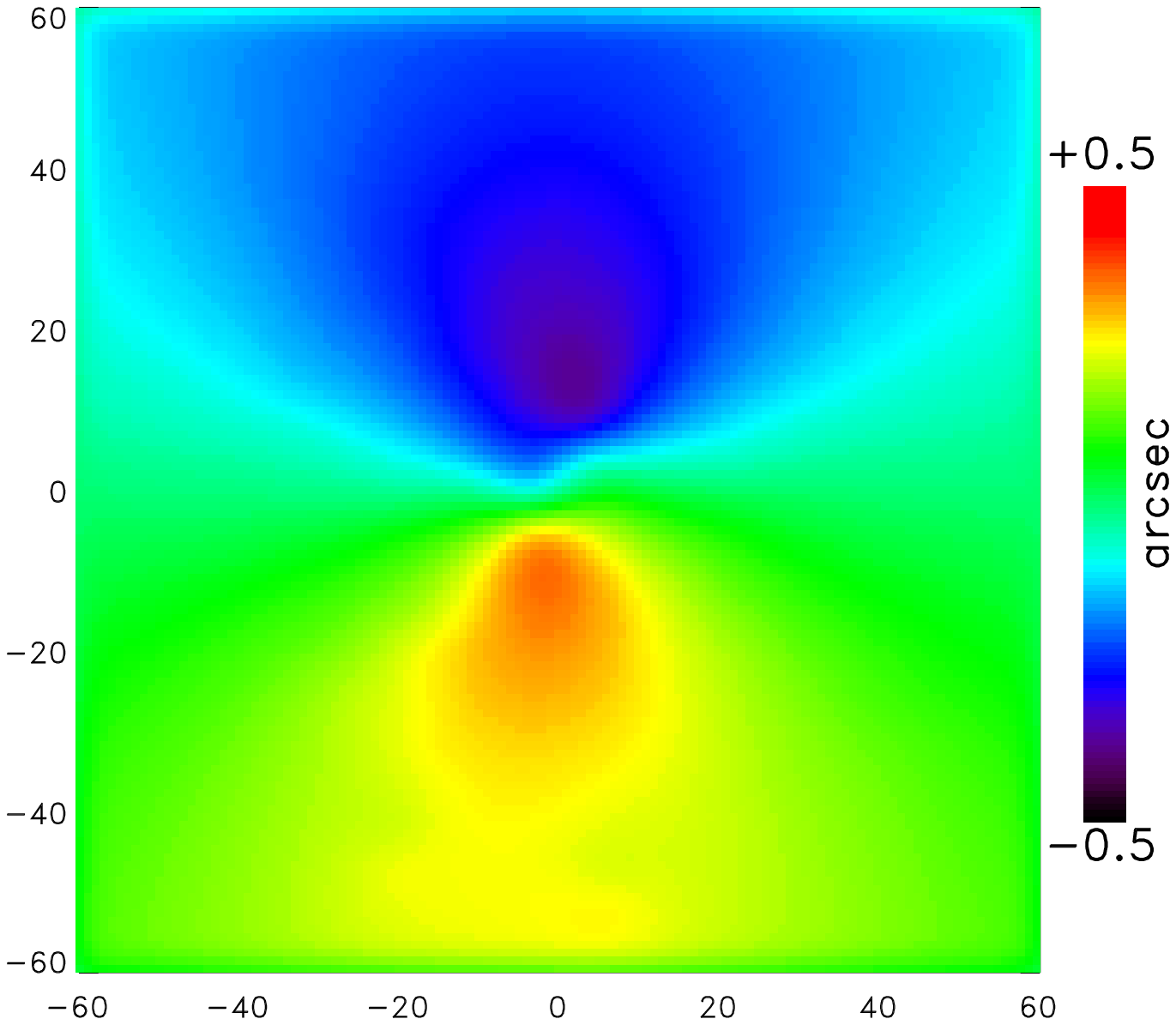}
    }
  }
  \vspace{9pt}
\caption{The signals shown in Figure~\Rf{txty} smoothed
with a $5$ arcsec beam.}
\label{fig:gauss}
\end{figure}

A realistic CMB experiment will have a finite width beam. Figure~\Rf{gauss}
shows the results of smoothing the signal with a beam of full width half maximum of 
$5''$. Even after such smoothing, 
there remains a noticeable difference between the clumpy and smooth galaxies.


\section{Conclusion}

Mass surrounding a galaxy lenses the background cosmic photons
arriving from the surface of last scattering at $z\simeq 1100$. 
This lensing can be used to probe the dark matter distribution around
galaxies.  On the relevant scales -- of order an arcminute --
 the background is nearly a dipole, and the 
amplitude of the signal is of order $1''\times \partial T/\partial\theta \sim 0.2\mu$K.
This signal can be used in two fundamentally different ways.
On the one hand, one can average over many foreground galaxies so as
to obtain a measure of the galaxy-mass correlation function.
We have shown that given enough sky coverage a CMB experiment with an arcminute beam size
and noise of $10\mu$K per pixel would be able to make
a measurement of this correlation function that is competitive with SDSS.
Instruments like the Penn Bolometric Array -- a $3$mm camera
for the $100$m Green Bank telescope -- hold some promise for 
beginning a more detailed exploration of galaxy mass distributions.
It is worth noting that, unlike measurements of lensing-induced
shear or amplification, which are sensitive to the second derivative 
of the lens gravitational potential, $\Phi$, this proposed measurement of 
the deflection angle is sensitive to $\nabla\Phi$.  In principle,
the inversion to obtain $\Phi$, is far easier from $\nabla\Phi$ than 
from $\nabla\nabla\Phi$."

In the distant future, we might be able to go further and probe 
the dark matter halo of an individual galaxy.
For noise levels reached today, the signal to noise from
a single galaxy is of order $1/100$. Either noise levels will
have to come down appreciably, 
or a different source needs to be used. 
This latter option is intriguing.
One particularly appealing possibility is 
to use the gradient caused by the thermal Sunyaev-Zel'dovich (SZ) effect (the scattering of CMB
photons by electrons in an ionized plasma) from a background galaxy cluster. 
This results from CMB photons scattering  off hot electrons in the cluster's
ionized IGM.  A typical expected temperature gradient is of order $500\mu$K arcmin$^{-1}$, 
almost two orders of magnitude above that from the primordial CMB.
If a galaxy were positioned along the line of sight to a cluster, where
the SZ effect would look nearly like a gradient on sub-arcminute scales,
then we would expect the galaxy lensing signal to be $\sim 5-10 \mu$K.
This is then comparable to current noise levels in CMB experiments, so
might be the best hope of measuring this fascinating signal in the near future.

The work of SD is supported by the DOE, by NASA grant NAG5-10842, 
and by NSF Grant PHY-0079251. GDS is supported by a DOE  grant to 
astrophysics theory group  at CWRU.  GDS thanks the astrophysics group at Fermilab
where this work began and the Kavli Institute for Theoretical Physics
where much of this work was completed. The Kavli Institute is supported in part by National Science
Foundation grant PHY99-07949. We thank Asantha Cooray and Andrey Kravstov for very helpful
comments on an earlier draft and Lam Hui and Charles Keeton for their insight and advice.

\newcommand\spr[3]{{\it Physics Reports} {\bf #1}, #2 (#3)}
\newcommand\astron[3]{{\it Astronomical J.} {\bf #1}, #2 (#3)}
\newcommand\saa[3]{{\it Astronom and Astrophys.} {\bf #1}, #2 (#3)}
\newcommand\sapj[3]{ {\it Astrophys. J.} {\bf #1}, #2 (#3) }
\newcommand\sprd[3]{ {\it Phys. Rev. D} {\bf #1}, #2 (#3) }
\newcommand\sprl[3]{ {\it Phys. Rev. Letters} {\bf #1}, #2 (#3) }
\newcommand\np[3]{ {\it Nucl.~Phys. B} {\bf #1}, #2 (#3) }
\newcommand\smnras[3]{{\it Monthly Notices of Royal
	Astronomical Society} {\bf #1}, #2 (#3)}
\newcommand\splb[3]{{\it Physics Letters} {\bf B#1}, #2 (#3)}
\newcommand\astroph[1]{{\tt astro-ph/}{#1}}


\begin{thebibliography}{99}


\bibitem{moore} B.~Moore et al., \sapj{524}{L19}{1999}

\bibitem{andrei} A.~Klypin, A.~V.~Kravtsov, O.~Valenzuela, and
	F.~Prada, \sapj{522}{82}{1999}


\bibitem{bullock} J.~S.~Bullock, A.~V.~Kravtsov, and D.~H.~Weinberg,
	\sapj{539}{517}{2000}

\bibitem{spergel} D.~N.~Spergel and P.~J.~Steinhardt, \sprl{84}{3760}{2000}

\bibitem{yoshida} N.~Yoshida, V.~Springel, S.~D.~M.~White, and G.~Tormen, \sapj{544}{L87}{2000}

\bibitem{colin} P.~Colin, V.~Avila-Reese, O.~Valenzuela, and C.~Firmiani, \sapj{581}{777}{2002}

\bibitem{burkert} E.~D'Onghia and A.~Burkert, \sapj{586}{12}{2003}
 
\bibitem{mao} S.~Mao and P.~Schneider, \smnras{295}{587}{1998}

\bibitem{metcalfe} R.~B.~Metcalfe and P.~Madau, \sapj{563}{9}{2001}

\bibitem{dalal} N.~Dalal and C.~S.~Kochanek, \sapj{572}{25}{2002}

\bibitem{chiba} M.~Chiba, \sapj{565}{17}{2002}

\bibitem{keeton} C.~Keeton, astro-ph/0111595

\bibitem{tyson} J.~A.~Tyson, F.~Valdes, J.~F.~Jarvis, and A.~P.~Mills,
\sapj{281}{L59}{1984}

\bibitem{brainerd} T.~G.~Brainerd, R.~D.~Blanford, and I.~Smail,
\sapj{466}{623}{1996}

\bibitem{fischer} P.~Fischer et al., \astron{120}{1198}{2000}

\bibitem{smith} D.~Smith, G.~Bernstein, P.~Fischer, and M.~Jarvis,
\sapj{551}{643}{2000}

\bibitem{wkl} G.~Wilson, N.~Kaiser, G.~A.~Luppino, and L.~L.~Cowie,
\sapj{555}{572}{2001}


\bibitem{mckay} T.~A.~McKay et al., \astroph{0108013}

\bibitem{song} Y.-S.~Song, A.~Cooray, L.~Knox, and M.~Zaldarriaga,
\astroph{0209001}

\bibitem{hoekstra} H.~Hoekstra et al., \astroph{0206103}

\bibitem{klein} M.~Kleinheinrich et al., \astroph{0304208}

\bibitem{berwei} A.~A.~Berlind and D.~H.~Weinberg, \sapj{575}{587}{2002}

\bibitem{guzik} J.~Guzik and U.~Seljak, \smnras{321}{439}{2001}

\bibitem{cooray} A.~Cooray, \astroph{0206068}

\bibitem{weidav} D.~H.~Weinberg, R.~Dav'e, N.~Katz, and L.~Hernquist,
\astroph{0212356}

\bibitem{bwb} A.~A.~Berlind et al., \astroph{0212357}

\bibitem{jain} B.~Jain, R.~Scranton, and R~K.~Sheth, \astroph{0304203}

\bibitem{sz} U.~Seljak and M.~Zaldarriaga, \sapj{538}{57}{2000}

\bibitem{mc} S.~Dodelson, {\it Modern Cosmology} (Academic Press, Amsterdam, 2003)

\bibitem{peiris} H.~V.~Peiris and D.~N.~Spergel, \sapj{540}{605}{2000}

\bibitem{wmap} C.~L.~Bennett et al., \astroph{0302207}

\bibitem{mcex} See, e.g., J.~A.~Peacock, {\it Cosmological Physics} (Cambridge University
Press, Cambridge, 1999) or Ref.~\cite{mc}, Chapter 10.

\bibitem{nfw} J.~F.~Navarro, C.~S.~Frenk, and S.~D.~M.~White,
	\sapj{462}{563}{1996}

\bibitem{bryan} G.~Bryan and M.~Norman, \sapj{495}{80}{1998}

\bibitem{bul337} J.~Bullock et al., \smnras{321}{559}{2001}

\bibitem{hirata} C.~Hirata and U.~Seljak, \sprd{67}{043001}{2003}


\bibitem{tegmark} M.~Tegmark and G.~Efstathiou, \smnras{281}{1297}{1996}

\bibitem{toffolatti} M.~Toffolatti et al., \smnras{297}{117}{1998}

\bibitem{teho} M.~Tegmark, D.~J.~Eisenstein, W.~Hu, and A.~de Oliveira-Costa, 
\sapj{530}{133}{2000}

\bibitem{wmapfor} C.~L.~Bennett et al., \astroph{0302208}

\bibitem{zhang} P.~J.~Zhang, U.-L.~Pen, and B.~Wang, \sapj{577}{555}{2002}

\bibitem{bond} J.~R.~Bond, M.~I.~Ruetalo, J.~W.~Wadsley, and M.~D.~Gladders,
\astroph{0112499}

\bibitem{hernquist} V.~Springel, M.~White, and L.~Hernquist,
\sapj{549}{681}{2001};
M.~White, L.~Hernquist, and V.~Springel,
\sapj{579}{16}{2002}
\bibitem{wmapspe} D.~N.~Spergel et al., \astroph{0302209}


\bibitem{vishniac} J.~P.~Ostriker and E.~Vishniac, \sapj{306}{51}{1986};
E.~Vishniac, \sapj{322}{597}{1987}

\bibitem{hu} W.~Hu, \sapj{529}{12}{2000}

\bibitem{sil1} A.~C.~da Silva, et al. \smnras{326}{155}{2001}

\bibitem{sil2} A.~C.~da Silva, et al., \sapj{561}{L15}{2001}

\bibitem{gnedin} N.~Y.~Gnedin and A.~Jaffe, \sapj{551}{3}{2001}

\bibitem{val} P.~Valageas, A.~Balbi, and J.~Silk, \saa{367}{1}{2001}

\bibitem{ma} C.-P.~Ma and J.~N.~Fry, \sprl{88}{211301}{2002}

\bibitem{zh2} P.~Zhang, U-L.~Pen, and H.~Trac, \astroph{0304534}

\bibitem{aghanim} N.~Aghanim, S.~Prunet, O.~Forni, and F.~R.~Bouchet,
\saa{334}{409}{1998}

\bibitem{coo2} A.~Cooray, \sprd{65}{083518}{2002}

\bibitem{cooche} A.~Cooray and X.~Chen, \sapj{573}{43}{2002}

\bibitem{chluba} J.~Chluba and K.~Mannheim, \astroph{0208392}

\bibitem{pkeet} C.~Keeton, private communication


\end{thebibliography}
\end{document}